# MoS$_2$ Dual-Gate MOSFET with Atomic-Layer-Deposited Al$_2$O$_3$ as Top-Gate Dielectric

Han Liu and Peide D. Ye, *Senior Member, IEEE*

*Abstract* - We demonstrate atomic-layer-deposited (ALD) high-k dielectric integration on two-dimensional (2D) layer-structured molybdenum disulfide (MoS$_2$) crystals and MoS$_2$ dual-gate n-channel MOSFETs with ALD Al$_2$O$_3$ as top-gate dielectric. Our C-V study of MOSFET structures shows good interface between 2D MoS$_2$ crystal and ALD Al$_2$O$_3$. Maximum drain currents using back-gates and top-gates are measured to be 7.07mA/mm and 6.42mA/mm at $V_{ds}$=2V with a channel width of 3 μm, a channel length of 9 μm, and a top-gate length of 3 μm. We achieve the highest field-effect mobility of electrons using back-gate control to be 517 cm$^2$/Vs. The highest current on/off ratio is over 10$^8$.

## I. INTRODUCTION

EVER since the advent of graphene in 2004 [1], the electronic properties of 2D layered-structure materials have been intensively investigated since their thickness can be pushed down to a few nanometers or even less, where a series of novel physical, chemical, and mechanical properties are observed. The layer-structured material family includes graphene, boron nitride (BN), MoS$_2$, topological insulators such as Bi$_2$Te$_3$, Bi$_2$Se$_3$, and many others [1-5]. As Moore's Law is approaching its physical limit, an alternative material is urgently needed as a substitute for future logic transistor applications [6,7]. Although graphene has been widely believed as a promising candidate, its gapless nature limits its potential application as logic transistors [8,9]. However, in great contrast to graphene, MoS$_2$ enjoys its uniqueness in the device applications due to its semiconductor-like bandgap [5]. Also, Due to the nature of its layered-structure, single atomic layer MoS$_2$ transistors with ultrathin body channel has the advantage in nanometer-scale MOSFETs of being immune to the short-channel effects, compared to the state-of-art SOI counterparts which are greatly limited by short-channel effects. Research in MoS$_2$ electronics is still in its infancy. The first experimentally demonstrated single-layer MoS$_2$ transistor has already been shown to have a mobility of over 200 cm$^2$/Vs, a sub-threshold swing (SS) of 74 mV/decade and on/off ratio of ~10$^8$ [3]. Following NEGF simulations theoretically predicted the perfect performance limits of MoS$_2$ thin film transistors [10]. In this Letter, we focus on the integration of ALD high-k oxides on this kind of layered-structure with potentially chemical inert surface. We demonstrate a top-gated Al$_2$O$_3$/MoS$_2$ MOSFET with electron mobility of 517 cm$^2$/Vs and on-off ratio of 10$^8$.

Han Liu and Peide D. Ye are with the School of Electrical and Computer Engineering and the Birck Nanotechnology Center, Purdue University, West Lafayette, IN 47907 USA (Tel: +1 (765) 494-7611, Fax: +1 (765) 496-7443, E-mail: yep@purdue.edu).

## II. EXPERIMENT

MoS$_2$ thin flakes were mechanically exfoliated by the classical scotch-tape technique and then transferred to a heavily doped Si substrate capped with 300 nm SiO$_2$. Al$_2$O$_3$ was deposited on MoS$_2$ flakes within an ASM F120 ALD reactor. Trimethylaluminum (TMA) and water were used as precursors at a temperature of between 200 ℃ and 400 ℃ with 111 cycles, which yields approximately ~10nm Al$_2$O$_3$ for an ideal ALD process. A dimension 3100 AFM system was used to examine the surface morphology before and after ALD deposition. MOSFET fabrication was then performed after identifying the appropriate ALD process window. A 16 nm ALD Al$_2$O$_3$ was deposited on MoS$_2$ flakes under 200 ℃ growth temperature. After Al$_2$O$_3$ growth, source and drain regions were defined using optical lithography with a spacing of 9μm. A 20/50 nm Ni/Au was deposited as the source/drain contacts and Ti/Au was used for gate. The gate length is ~3μm with ~3 μm spacings to source/drain.

## III. RESULTS AND DISCUSSION

Figure 1(a) and 1(b) shows the representative AFM images of MoS$_2$ surface morphology after 111 cycles of Al$_2$O$_3$ deposition under 200 ℃ and 400 ℃, respectively. In the absence of intensive research of ALD growth on 2D electronic materials, we can simply take the graphene as a reference. Earlier studies have shown that direct Al$_2$O$_3$ growth by TMA and water is not possible on the graphene basal plane, while growth occurs only at the graphene edges. This is understood by the fact that dangling bonds only exist in graphene edges but not on the basal plane [11,12]. However, for MoS$_2$, it is obvious that the ALD growth is easier than that on graphene. From Figure 1(a), we can see that at 200 ℃, the Al$_2$O$_3$ thin film is visually uniform. In addition, the step height between the flake and SiO$_2$ substrate is around 8 nm, which is similar to that before ALD growth. At the elevated temperature of 400 ℃, though we still observe continuous ALD growth at flake edges as shown in Figure 1(b), the step height has disappeared as Al$_2$O$_3$ is growing only on SiO$_2$, while only small areas of Al$_2$O$_3$ are formed on the MoS$_2$ basal plane. This temperature sensitivity observed in Al$_2$O$_3$ deposition indicates that the growth mechanism on MoS$_2$ mostly relies on physical absorption on the basal plane during the initial growth stage, whereas desorption is greatly enhanced at higher temperature. At the step edges of the layers, stronger chemical bonds between MoS$_2$ and ALD precursors are formed due to dangling bonds, thus even at higher temperature, the ALD growth on step edges is still uniform and continuous. 300 ℃ grown Al$_2$O$_3$

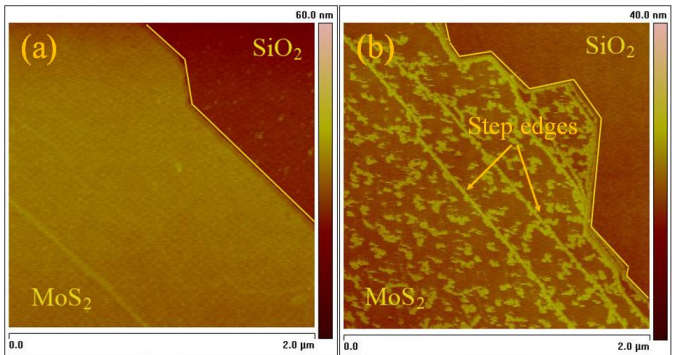

**Fig. 1** AFM images showing MoS$_2$ flakes with 111 cycles of Al$_2$O$_3$ grown at (a) 200 °C and (b) 400 °C.

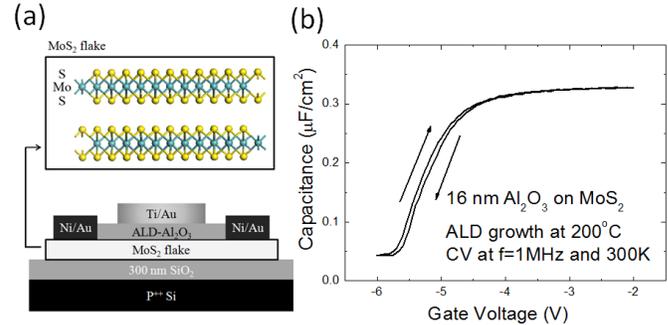

**Fig. 2** (a) Device structure of MoS$_2$ dual gate MOSFET, (b) 1 MHz high-frequency C-V characteristic of the MOSFET device measured at room temperature in darkness.

films on MoS$_2$ also show un-uniformity and poor electrical insulating properties. The ALD window for 2D layered-structure materials is significantly reduced, compared to that of bulk materials, such as Si, Ge and III-V. Therefore, the ALD process must be carefully optimized to simultaneously achieve geometrical uniformity and good electrical properties (high dielectric constant, large electrical strength, low gate leakage current, etc.).

The MoS$_2$ MOSFET was fabricated on a ~15 nm thick flake which contains about 23 MoS$_2$ monolayers. Final device structure is shown in Figure 2(a). We didn't reduce the flake thickness to a monolayer since the bandgap of ultrathin MoS$_2$ crystal increases and could become 1.8 eV for monolayer [5]. C-V measurement is carried out in order to evaluate the interface quality between ALD Al$_2$O$_3$ and MoS$_2$ crystals, as shown in Figure 2(b). Source and drain are grounded, while a voltage bias on the top-gate is applied. The area of the capacitor is only ~12 μm$^2$, making the low frequency C-V curve rather noisy (not shown). The high frequency C-V (with hysteresis) at 1 MHz shows a clear transition from accumulation to depletion for a typical n-type MOS capacitor. A moderate hysteresis of ~80 mV exhibits in the curves, showing that the ALD Al$_2$O$_3$ film grown at 200 °C on MoS$_2$ and the interface are both of good quality.

Figure 3(a) and 3(b) show the transfer characteristics and transconductance of the device from both the top-gate and the back-gate. The charge neutrality level (CNL) of MoS$_2$ is located slightly under the conduction band, thus making it easy for an accumulation-type nMOSFET [13,14]. The transfer characteristics of the top-gate suffer from a very large negative threshold voltage ($V_{th}$) shift, as attributed to the existence of large amount of positive fixed charges in the bulk oxide, due to the comparatively lower deposition temperature [15]. The leakage current is also measured in the same device, and is less than $2 \times 10^{-4}$ A/cm$^2$ in the measurement range of -6V to +3V. The highest drain current density achieved at $V_{ds}$=1V using back-gate modulation is 3.07 mA/mm and an on/off ratio greater than $10^8$. This superior on/off ratio compared to graphene exists because of the 1.2 eV bandgap. The greatest current density from top-gate is about 2 orders of magnitude smaller than that from the back-gate. This big difference comes from the non-self-aligned top-gate device structure. From Figure 2(a) we can see that the heavily doped Si substrate has a "global" control over the entire flake. With increasing back-gate voltage, the carriers in the MoS$_2$ flake are accumulated and thus the contact resistance between the Ni/Au source/drain and the MoS$_2$ flake would be reduced, as the flake is being heavily "doped" by the electric field, while the top-gate can only modulate part of the channel underneath the top-gate. The peak extrinsic transconductance ($g_m$) from back-gate control is 0.165 mS/mm at $V_{ds}$=1V. We can extract the field-effect mobility to be 517 cm$^2$/Vs, which is a factor of 2.6 larger than the reported value in Ref. [3]. Since the significant contact resistance is not subtracted, the intrinsic field mobility of MoS$_2$ channel is even larger. The extrinsic $g_m$ from top-gate modulation is 0.61 mS/mm, smaller than the back-gate $g_m$, due to the large contact resistance and access resistance when back-gate is floating, which is discussed below. The hysteresis of the top-gate transfer curves is much smaller than that of the back-gate curves, similar to the C-V curves. The subthreshold swing (SS) for top-gate is ~140 mV/dec at $V_{ds}$=1V. The interface trap density is estimated to be $2.4 \times 10^{12}$/cm$^2$-eV at MoS$_2$ and ALD Al$_2$O$_3$ interface and would be further reduced by optimizing the process. Considering that there is minimal process refinement, such as no surface passivation and the low ALD growth temperature, this may imply the interface states issue between 2D crystals and ALD high-k dielectrics is very forgiving.

Figure 4(a) and 4(b) shows the drain current versus drain voltage under a variety of back- and top-gate biases. The gate biases range from 50 V to 20 V with a -5 V step for back-gate, and from -3 V to -6 V with a -0.5V step for top-gate. For the top-gate measurement, a back-gate voltage of 50 V is applied to reduce the contact resistance and access resistance. Consequently, the maximum current density for top-gate modulation has now been increased to 6.42 mA/mm, back to the same level of that from back-gate modulation. The maximum extrinsic $g_m$ at $V_{ds}$=2 V are 0.318 mS/mm and 2.83 mS/mm for back-gate and top-gate modulation, respectively. By roughly extracting the contact resistance, the quasi-intrinsic field-effect mobility from the top-gate device is increased to 125 cm$^2$/Vs. The big difference between field-effect mobility from top- and from back-gated devices could be ascribed the different interface conditions between the MoS$_2$/SiO$_2$ interface and MoS$_2$/Al$_2$O$_3$ interface. For the former, after flake is transferred to the substrate and is in physical contact, the interface remains intact throughout the fabrication. However, the top interface is strongly correlated with the ALD process, similar to the challenges found in our previous work on the topological insulator Bi$_2$Te$_3$. The physical absorption process during the initial ALD growth of high-k dielectrics still needs a comprehensive study in order to develop a deep understanding of the ALD growth mechanism on these novel 2D electronic materials. The definition of the "interface states" at these 2D atomic crystal/Al$_2$O$_3$ interfaces needs to be revised since there are no traditional dangling bonds at these interfaces at all.

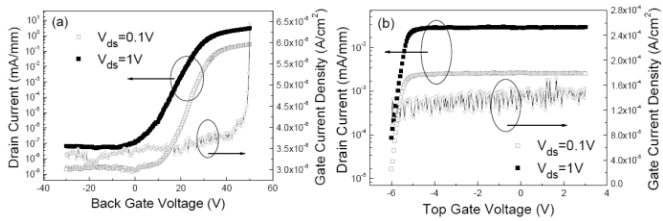 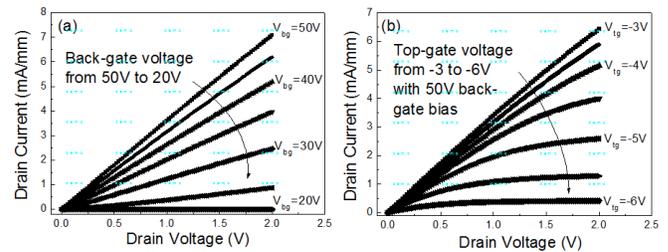

**Fig. 3** Transfer characteristics of the MoS$_2$ dual gate MOSFET from (a) back-gate and (b) top-gate controls. The back- and top-gate leakage current density from 300 nm SiO$_2$ and 16 nm Al$_2$O$_3$ are also presented.

**Fig. 4** I$_d$-V$_d$ characteristics of MoS$_2$ dual gate MOSFET with back-gate voltage stepped from 50 V to 20 V. (b) I$_d$-V$_d$ characteristics of MoS$_2$ dual gate MOSFET with top gate voltage stepped from -3 V to -6 V while a back-gate voltage of 50 V is applied.

## IV. Conclusion

In summary, we have experimentally demonstrated MoS$_2$ MOSFET with ALD Al$_2$O$_3$ as top-gate dielectric. AFM, C-V and I-V studies show that ALD high-k dielectrics can be directly deposited on MoS$_2$ at low growth temperatures and the MoS$_2$/Al$_2$O$_3$ interface is of good quality. The high electron field mobility, good subthreshold swing, and excellent drain current on/off ratio are demonstrated on the fabricated MoS$_2$ nMOSFET.